\documentclass[%
 reprint,
 amsmath,amssymb,
 aps,
]{revtex4-2}

\usepackage{graphicx}
\usepackage{dcolumn}
\usepackage{bm}
\usepackage{color}
\usepackage{natbib}

\begin{document}

\preprint{APS/123-QED}



\title{Initial states dependence of phase behaviors in dense active system}%





\author{Lu Chen\textsuperscript{1,2}}\email[Corresponding author. Email: ]{lu_chen@mail.bnu.edu.cn}
\author{Bokai Zhang\textsuperscript{3}}%
\author{Z. C. Tu\textsuperscript{1}}%

\affiliation{$^1$Department of Physics, Beijing Normal University, Beijing 100875, China}
\affiliation{$^2$Complex Systems Division, Beijing Computational Science Research Center, Beijing 100193, China}
\affiliation{$^3$School of Physical Science and Technology, Southwest University, Chongqing 400715, China}%

\date{\today}

\begin{abstract}
There are rich emergent phase behaviors in non-equilibrium active systems. Flocking and clustering are two representative dynamic phases. The relationship between these two phases is still unclear. In the paper, we numerically investigate the  evolution of flocking and clustering in a system consisting of self-propelled particles with active reorientation . We consider the interplay between flocking and clustering phases under different initial states, and observe an unstable domain in order parameters phase diagrams due to initial states even in the absence of an explicit attraction. This point is different from the previous finding that active angular fluctuations lead to an earlier breakdown of collective motion and the emergence of a new bi-stable regime in the aligned active particles~\cite{NJf073033}. In particular,  we find that the existence of bi-stable states is due to the diversity  of dynamic paths arising from different initial states.  By increasing (decreasing) the initial degree of ordering, the bi-stable state can be shifted to a more ordered flocking (disordered clustering) state. These results enlighten us pave the way to manipulate emergent behaviors and collective motions of active system.

\end{abstract}

\maketitle


\section{\label{sec:level1}INTRODUCTION }
Active systems have been the subjects of intense attention~\cite{PhysRevLett111190,pnas1521151113,PhysRevE101022602,PhysRevLett123098001}. The emergence of collective motion appeared in such systems inspires us to consider the dynamics in the evolution process of active system. However, unlike equilibrum system determined by the principle of minimum free energy, these systems are far from equilibrium. So far there is a lack of widely acknowledge theory to describe the steady-state phase behaviors since nonequilibrium evolution is usually dependent on their dynamic paths, which is an interesting property widespread existing in many living and inanimate systems~\cite{PhysRevE88053004,2006Physical,PhysRevB433245,PhysRevB59,PhysRevE103}.

In the past few decades, extensive work has been performed towards initial states in network, natural and social systems~\cite{acsmacromol1c,jinh_a_01767,PhysRevE105024304,jacs9b02004,ComputerMethodsinAppl,PhysRevE100052201}. A series of intriguing behaviors emerge along with different dynamic paths, and the evolution of the system exhibits rich diversity. It is also found that the processing routes or transition pathways give rise to different structures, states, and properties of materials~\cite{PhysRevE_103_022,JOP10,PhysRevLett_110,PhysRevE_99_05,AdvancesinCondensedMatterPhysics,PhysRevLett_77_2077,PhysRevE_102}. In particular, one of the main manifestation of history dependence is initial-state dependence~\cite{PhysRevLett89023}. In recent years, many studies have shown that initial velocity distributions can influence the formation of coherent structures~\cite{PHYS_PLASMAS}, nonequilibrium criticality of magnetic systems~\cite{Physics_of_the_SolidState} and alignment of block copolymer micro\cite{2005Influence}. In the work of R. Großmann et al.~\cite{NJf073033}, they propose a model of active Brownian particles (ABPs) with
velocity alignment. Starting from the description of hydrodynamic theory of the model, it is deduced that the instability of the steady state of system is determined by passive noise, active speed noise and active angular fluctuations. The system relaxes into different stable states by starting from different initial conditions, which make us to consider whether this result also exists in other active models. And in such active model, how does the path affect the evolution of the system is still unclear. In the meantime, there is few research specifically concerning the interplay between the collective motion and path-dependent clustering and flocking phases which is related to self-organization in the active system. Consequently, it is worth to understand deeply the mechanisms controlling the key characteristics of specific phase behaviors which are sensitive to the history.

In this work, we focus on the impact of initial states on the phase behaviors in the active system, which is consisted of self-propelled particles with  active reorientation. Previously, the rotations of self-propelled particles are common in active matter by assigning a torque~\cite{SoftMatterC9SM,SelfProRods} or adding angle transformations~\cite{2019Controlling,EPL1209}.  In our current work, we propose a more natural active rotation inspired by experiment, in which each colliding particles spontaneously reorient and make it easier to separate. The coupling between self-propulsion and active reorientation produces richer dynamics compared with previous models merely involving translation or rotation. As a result, these particles may collectively aggregate to dynamic clusters or self-organize into alignment. By exploring evolution processes under different parameters, we reveal the complementary nature of clustering and flocking. We further study history-dependent property in the active system, and explore how the initial states reshape dynamic path and then affect the steady state of the system. We observe the discrepancy of steady states of the flocking and clustering phase under different initial states. The phase boundary is considered to comes from the competition between active reorientation and noise.  It is sensitive to initial particle position and velocity distribution. Interesting, the system could evolve into one of two nonequilibrium stable states by manipulating initial states. All these findings offer a new way to control collective behaviors by adjusting dynamic paths. 
\section{DYNAMIC MODEL}
 	\begin{figure}[h]
      \includegraphics[width=0.98\linewidth]{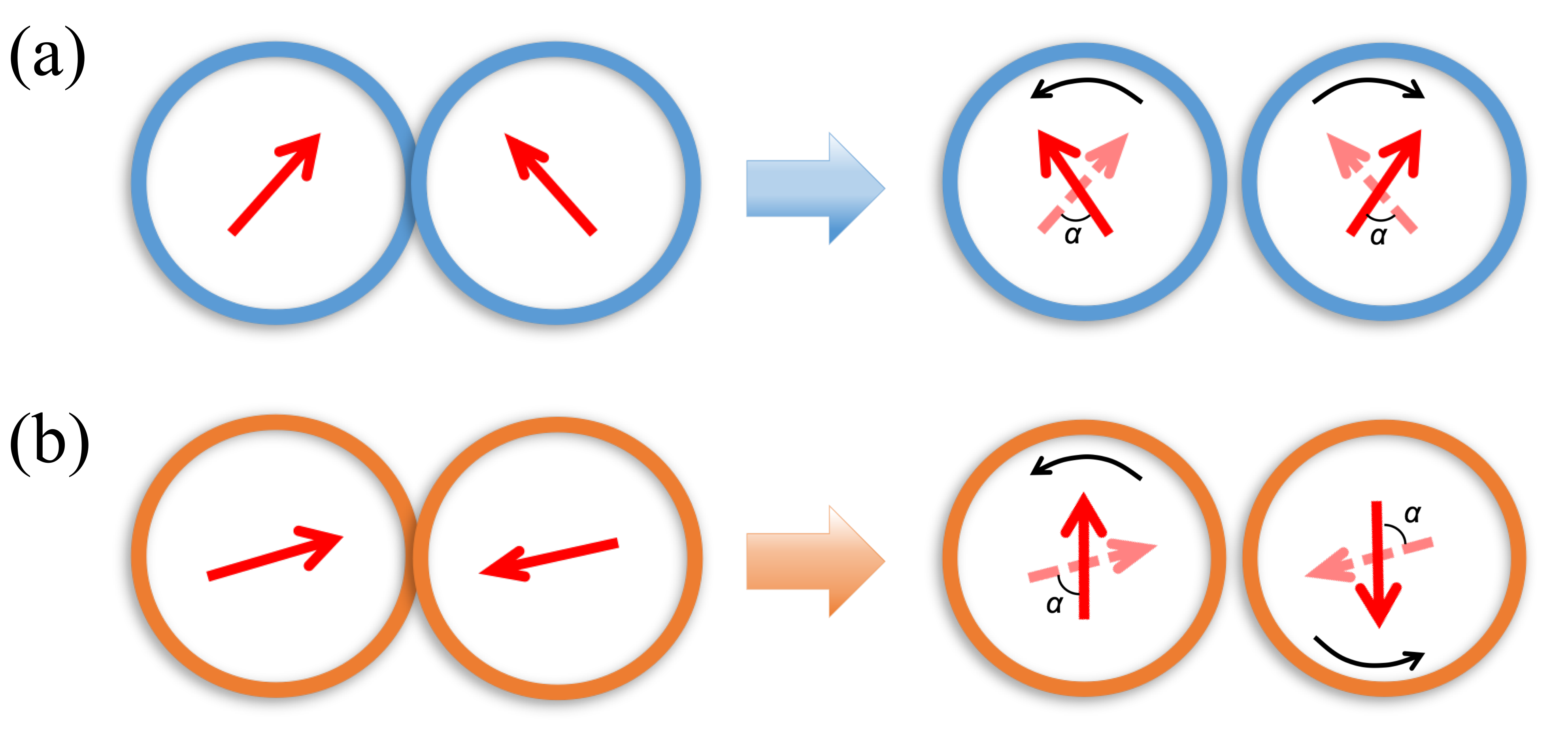}%
		\caption{(color online) (a-b) Schematic diagram of collision avoidance by active reorientation.
		}
		\label{fig1}
    \end{figure} 
    
In previous studies of active models, many emergent behaviors can be well-characterized by imposing explicit local alignment interactions~\cite{vicsek1995novel}, an inelastic collision rule~\cite{grossman2008emergence}, or a separation of time scales between orientation and propulsion~\cite{lam2015self,lam2015polar}. However, none of these models explicitly include collision avoidance in their rules of interaction, a strategy commonly adopted by motile animals in a herd~\cite{moussaid2011simple,chan2013collision}. Here we consider an active collision avoidance model consisting of self-propelled disks with a diameter $d$ in a plane. $\boldsymbol{r}_{i}$ and $\hat{\bm{n_i}}\equiv\left(\cos \theta_{i},\sin \theta_{i}\right)$ represent the position and self-propulsion direction of disk i (i = 1, . . . , N).
The dynamics of the ith disk is determined by the following equations:
\begin{eqnarray}
\Delta{\bm{r}}_{i}=\Delta{s} \bm{n}_{i} +\sum_{j=1,j\neq i}^{N}k\varepsilon_{ij} \left(r_{ij}-d\right) \frac{\bm{r}_{j}-\bm{r}_{i}}{{r}_{ij}} \Delta{t},\label{m1}\\ 
\Delta{\theta}_{i}=\eta_i\Delta{t}+ \sum_{j=1,j\neq i}^{N}\alpha\varepsilon_{ij} \left(\frac{\bm{r}_{j}-\bm{r}_{i}}{r_{ij}}\times\bm{n}_{i}\right) \cdot \bm{\hat{z}} .\label{m2}
\end{eqnarray}

The first equation describes translation of disk i. The first term $\Delta{s} \bm{n}_{i}$ represents the self-propulsion displacement along $\bm{n_i}$ in time-step $\Delta{t}$. The second term represents the contribution of collision when two disks come close to each other. $r_{ij}=|\bm{r}_{ij}|=|\bm{r}_{j}-\bm{r}_{i}|$ is the distance between the centers of disk i and disk j. If $r_{ij}<d$, then $\varepsilon_{ij}=1$, else $\varepsilon_{ij}=0$. Parameter $k$ represents the strength of two-body interaction between particles. The second equation describes rotation of disk i. When disk i is free, it rotates due to Gaussian white noise $\eta_i$  which satisfies $\left\langle\eta_{i}(t)\right\rangle=0$ and $\left\langle\eta_{i}(t) \eta_{j}\left(t^{\prime}\right)\right\rangle=\sigma\delta_{ij}\delta\left(t-t^{\prime}\right)$ where $\sigma$ is the strength of the noise. During collisions, the disks generate an active reorientation $\alpha$ in addition to the random Gaussian noise. This effect is reflected in the second term of equation~(\ref{m2}) where $\bm{\hat{z}}$ is a normal vector of the plane pointing the reader. We choose $\alpha>0$ in order to mitigate the collision between two disks in collision. This idea of effective collision avoidance is shown in Fig.\,\ref{fig1}. 

We numerically solve Eqs.~(\ref{m1}) and (\ref{m2}) using periodic boundary conditions. In our simulation, we set time-step $\Delta{t}=1$, displacement $\Delta s = 0.1d$, and packing fraction $\phi=0.3$. We have performed simulations with particle number $N = 100$ to 25,000 particles little difference is observed for systems with 1000 or more particles. Hence, we explore the kinetic properties of system by running simulation with 1000 disks.

\section{PHASE BEHAVIORS}
The dynamics of the system is controlled by three parameters: $\sigma$ of the Gaussian Noise, self-propulsion speed $\Delta s$ = 0.1d, and active reorientation step size $\alpha$. We have tested that the influence of self-propulsion speed on our system is mainly associated with the formation of cluster or living crystal states, such as the emergence of hexatic and solid order in~\cite{digregorio2018full}, which has little effect on dynamics. However, a certain amount of noise is necessary as well to induce a non-equilibrium phase separation. And the balance among noise, self-propulsion and particle interactions are is the key to demixing in an active Brownian particle~\cite{levis2014clustering}. So in this paper, we focus on the effects of noise and active reorientation combined on the phase dynamic behaviors of the system. In the mean time, if noise is big enough, the system goes into random states no matter how big or small active reorientation is. So next, we discuss the dynamics and interactions among these phases that appear in our system in detail.

\begin{figure}[h]
      \includegraphics[width=1.0\linewidth]{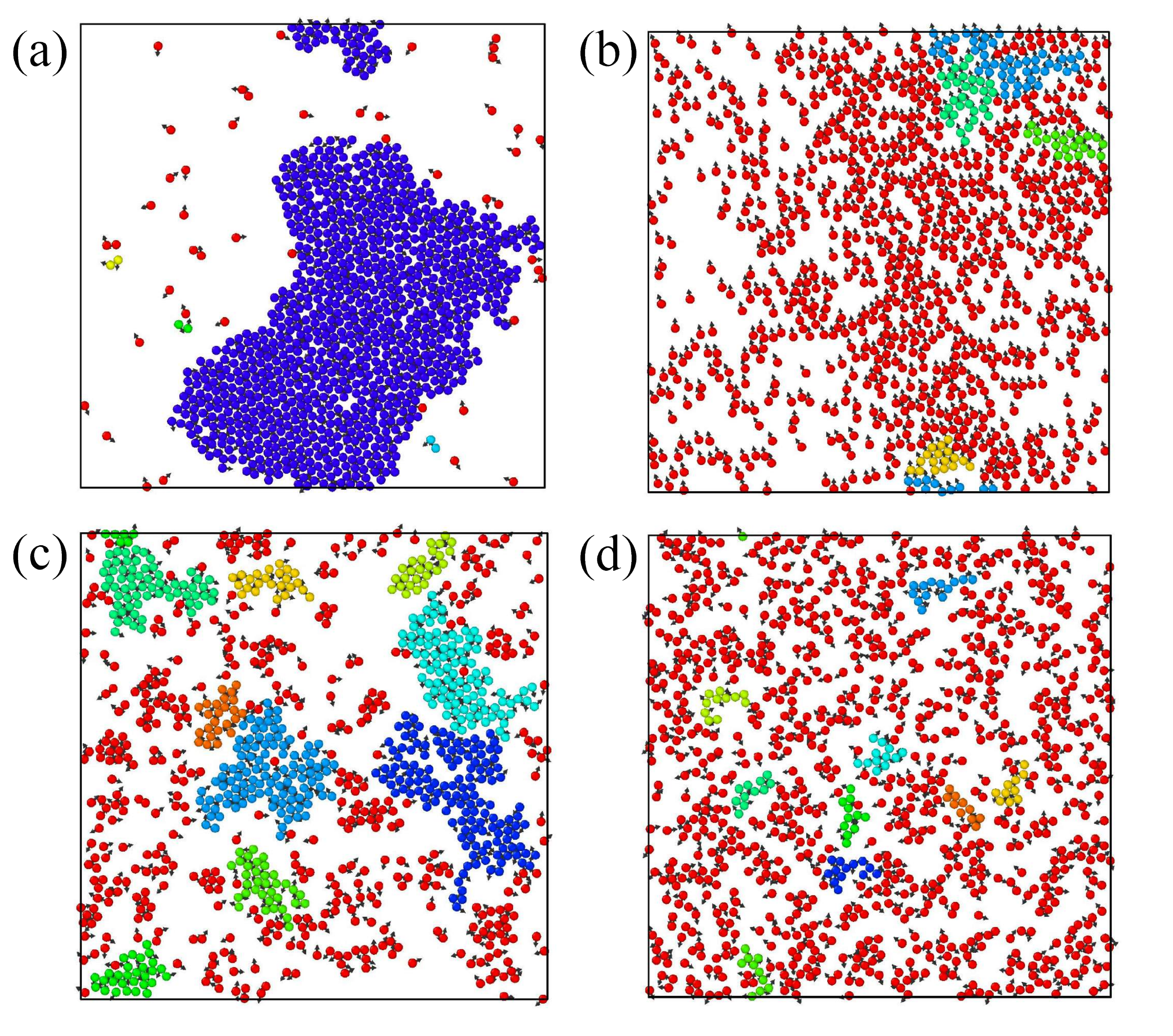}%
		\caption{(color online) Representative steady states: (a) clustering; (b) flocking; (c) partly disordered phase; (d) totally disordered phase. The arrows indicate the direction of particle velocity. Colors (except red) represent different clusters.		
		}
		\label{fig2}
    \end{figure}

For each given set of $\sigma$ and $\alpha$, we take random positions and directions of the particles as initial states to simulate the evolution of the system. Fig.\,\ref{fig2}(a-d) present some possible collective behaviors that arise from different evolutionary processes. As directions of particles evolve by actively reorienting themselves and rotating diffusion, which breaks the local balance and leads to the formation of cluster and related to the broken orientational symmetry.
As directions of particles evolve by actively reorienting themselves and rotating diffusely, the local balance and orientational symmetry of system are broken, which leads to the formation of cluster Fig.\,\ref{fig2}(a). Fig.\,\ref{fig2}(b) is a flocking state in which the orientations of all particles are approximate aligned while the position distribution is uniform. Fig.\,\ref{fig2}(c) and (d) represent two kinds of disordered phases. The first one includes small clusters even though the orientations of particles are random. The second one is totally disordered since the orientation and position of particles are all almost random.

\subsection{Order parameters}
\begin{figure}[t]
        \includegraphics[width=1.0\linewidth]{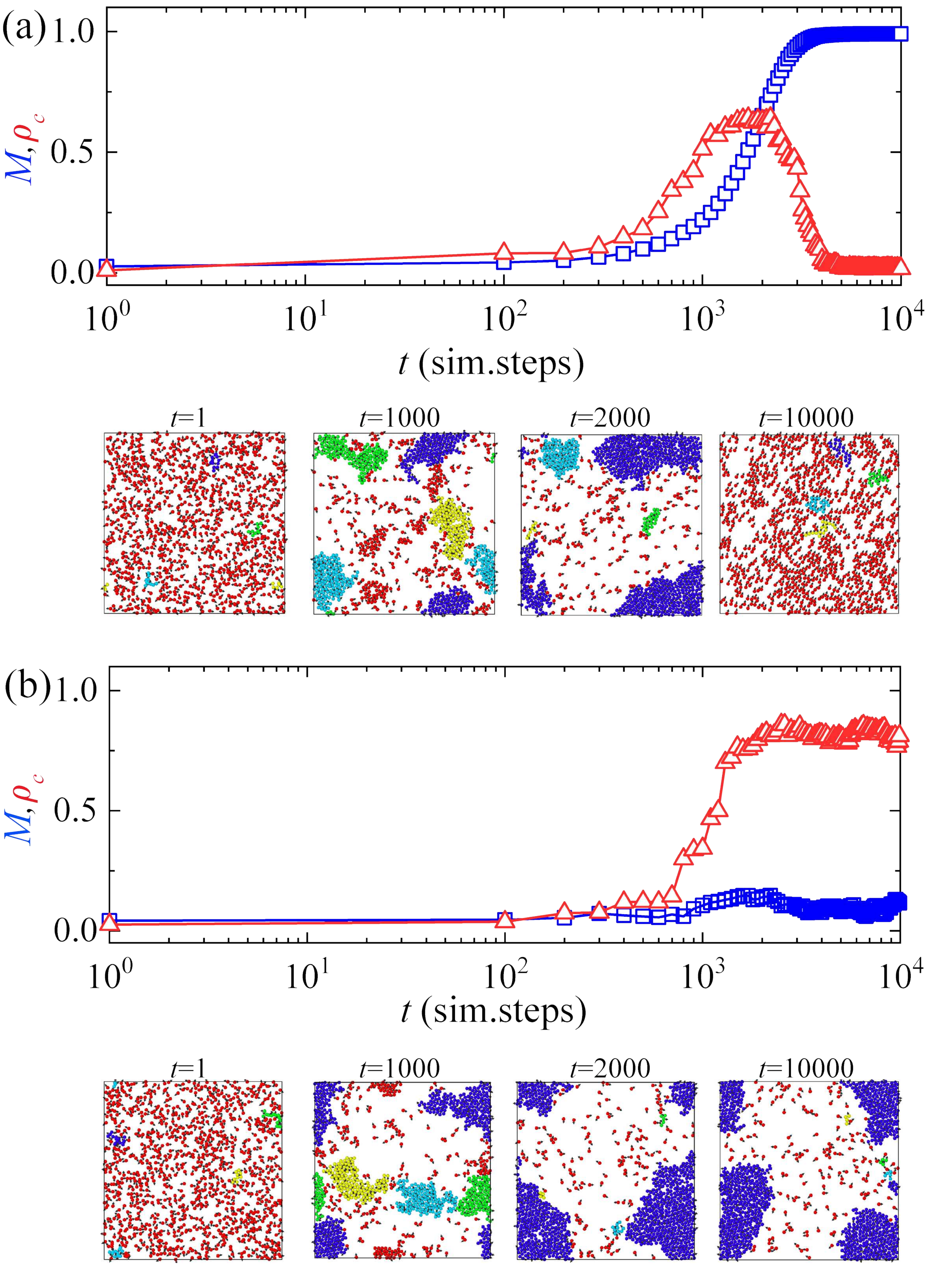}%
		\caption{(color online) Temporal evolution of order parameters and configurations. The system eventually evolved to (a) flocking state under $\alpha$ = 0.01, $\sigma$ = 0.01 and (b) clustering state under $\alpha$ = 0.01, $\sigma$ = 0.04.
		}
		\label{two-kinds-22-4-26}
    \end{figure}

We introduce two order parameters in order to characterize the steady states of system mentioned above. Flocking appeared in our system is described by an orientational order parameter,
\begin{equation}
M=\sqrt{\langle\cos \theta\rangle_{N}^{2}+\langle\sin \theta\rangle_{N}^{2}}
\end{equation}
 where the average $\langle\rangle_N$ is taken over all $N$ particles. $M\approx 1$ indicates perfect flocking phase and $M\approx 0$ disordered phase. To quantitatively identify the phase of system that separates into dense clustered and dilute gas-like state, we measure the local area density of each particle as~\cite{SoftMatter101039}: 
 \begin{equation}
   a_{l}=\frac{A}{A_{v}}
\end{equation}
 Here $A$ is the area of each disk, $A_{v}$ is the area of Voronoi cell of each disk. The larger size of cluster is, the more particles of larger $a_{l}$ there are. By numerical simulation, we found that when $a_{l}$ is greater than a certain threshold, the disk is located in the dense cluster. So in our work, we set the threshold as 0.7. And we introduce another order parameter $\rho_c$, the fraction of particles with $a_{l}>0.7$, to describe clustering. These two order parameters can give a good description of flocking and clustering, for example, in Fig.\,\ref{fig2}(a)-(d), $(M\approx0, \rho_c\approx0.92)$, $(M\approx1, \rho_c\approx0)$, $(M\approx0.7, \rho_c\approx0.26)$, $(M\approx0, \rho_c\approx0)$ respectively. In order to explore the dynamics of flocking and clustering, we analyzed a time series of $M$ and $\rho_c$. Fig.\,\ref{two-kinds-22-4-26}(a) shows time series under parameter $\alpha=0.01$ and $\sigma=0.01$. $M$ grows slowly first, then sharply increases and eventually reaches flocking state with $M$ close to 1. $\rho_c$ grows slowly at the beginning, then it reaches a transient plateau. Concurrent with the sharp increase of $M$, the transient plateau of $\rho_c$ collapses. The snapshots of the configurations are shown below Fig.\,\ref{two-kinds-22-4-26}(a). When the system starts from a disordered state, small condensation clusters appear in the system, and phase separation is triggered. Then the condensation nuclei come together and aggregate into a large cluster. Also, the non-equilibrium clustering appeared in the system were observed in other active models~\cite{peruani2006nonequilibrium,mccandlish2012spontaneous,chate2008modeling}. Next, as particles collide with each other, scattering with active reorientation competes with the self-diffusion process. As a result, the cluster gradually disintegrates and disperses. The final state of the system is highly flocking, similar to the flocking state in vicsek model~\cite{vicsekmodel}. In contrast, as shown in Fig.\,\ref{two-kinds-22-4-26}(b), under the parameter $\alpha=0.01,\sigma=0.04$, $\rho_c$ increases on a time scale similar to that in Fig.\,\ref{two-kinds-22-4-26}(a) and finally saturate around 0.8, whereas $M$ remains low value throughout the simulation. The evolution of configuration is relatively simpler. Particles interact with each other and gradually form stable clusters. There is no orientational order in the steady state. 

\subsection{Phase diagrams}
\begin{figure}[t]
\includegraphics[width=1.0\linewidth]{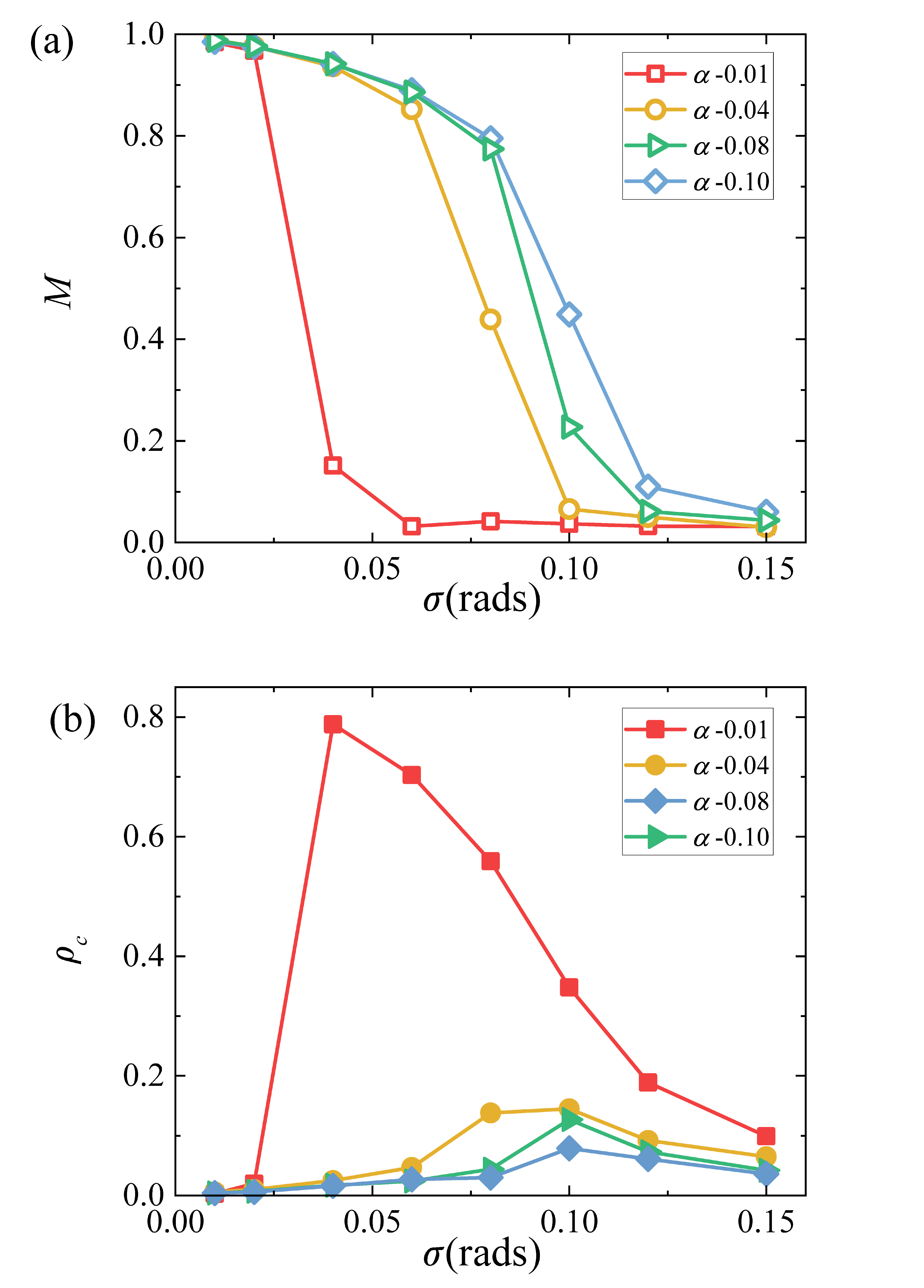}%
		\caption{(color online)The dependence of $M$ and $\rho_c$ on $\sigma$ with $\alpha=0.01,0.04,0.08,0.10$.
		}
		\label{m_roc_vs_diff_sig}
\end{figure}

The value of order parameters would fluctuate little for long enough time. In our model, the system keeps stable after a several thousand time-steps. So in order to characterize steady states, we measured the order parameters at 10,000 time-steps under different sets of $\{\alpha, \sigma\}$. We take random initial states and run simulations. We find that the system eventually evolves to be flocking state with $M \approx 1$ when the values of $\alpha$ and $\sigma$ are in certain region. But in the other region, the system can not reach the flocking state at all.
 In Fig.\,\ref{m_roc_vs_diff_sig}, we present the dependence of $M$ and $\rho_c$ on $\sigma$ with $\alpha=0.01,0.04,0.08,0.10$. As shown in In Fig.\,\ref{m_roc_vs_diff_sig}(a), when $\alpha$ is fixed, the system finally presents order or disorder state, which depends on the intensity of noise. With the increase of $\alpha$, the value of $\sigma$ at critical phase transition point also increases. As shown in In Fig.\,\ref{m_roc_vs_diff_sig}(b), similar behavior also manifests in the aggregation of clusters. In particular, the value of $\sigma$ at critical phase transition point for flocking and clustering is the same. 
 
\begin{figure}[t]
\includegraphics[width=1.0\linewidth]{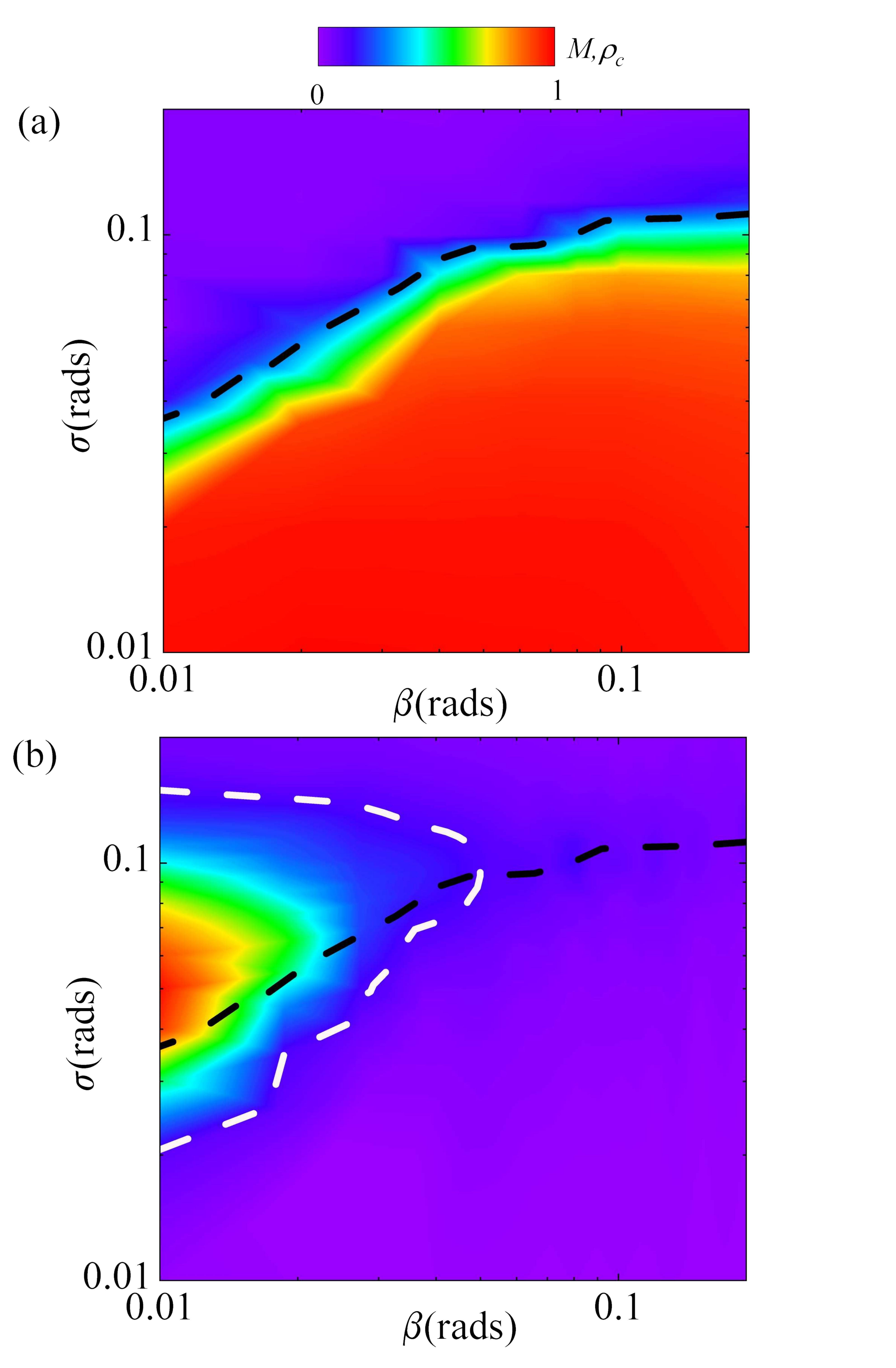}%
		\caption{(color online) Phase diagram of flocking and clustering. Contour map of flocking and clustering measured from simulations evaluated at the steady state from the initial conditions with a random distribution of particles (a-b).}
		\label{thre_domains_05_25_normal}
\end{figure}. 

 We draw the phase diagram according to the value of $M$ in the steady state in Fig.\,\ref{thre_domains_05_25_normal}(a). The dash line in the figure is the contour line of $M$ equals to around 0.3. When $\sigma$ is relative small, the system would reach a steady flocking state because active reorientation is dominant so that it promotes local velocity alignment and the formation of global flocking as in the domain below the black dash line. We also found both large active reorientation and random noise weaken the local velocity alignment, so that the system would be in disorder state as in the domain upon the dash line,  in which there is basically no global flocking. The phase diagram of $\rho_c$ is depicted in 
Fig.\,\ref{thre_domains_05_25_normal}(b), and the white dash line in the figure is the contour line of $\rho_c$ equals to around 0.1. And final clustering states are concentrated in the domain which surrounded by white lines. The overall phase behavior can be understood by considering the competition between the absorption rate of particles to clusters and escape rate of particles from clusters. When a particle collides with a cluster,  it has a probability either entering or escaping from the cluster due to active reorientation and noise. In the diagram of $\rho_c$, the regions with a high degree of clustering correspond to the regions where global alignment can not be formed in the diagram of $M$. High global flocking and large cluster can not coexist with each other.  

\section{THE EFFECT OF INITIAL STATES ON CLUSTERING AND FLOCKING}

\begin{figure}[t]
\includegraphics[width=1.0\linewidth]{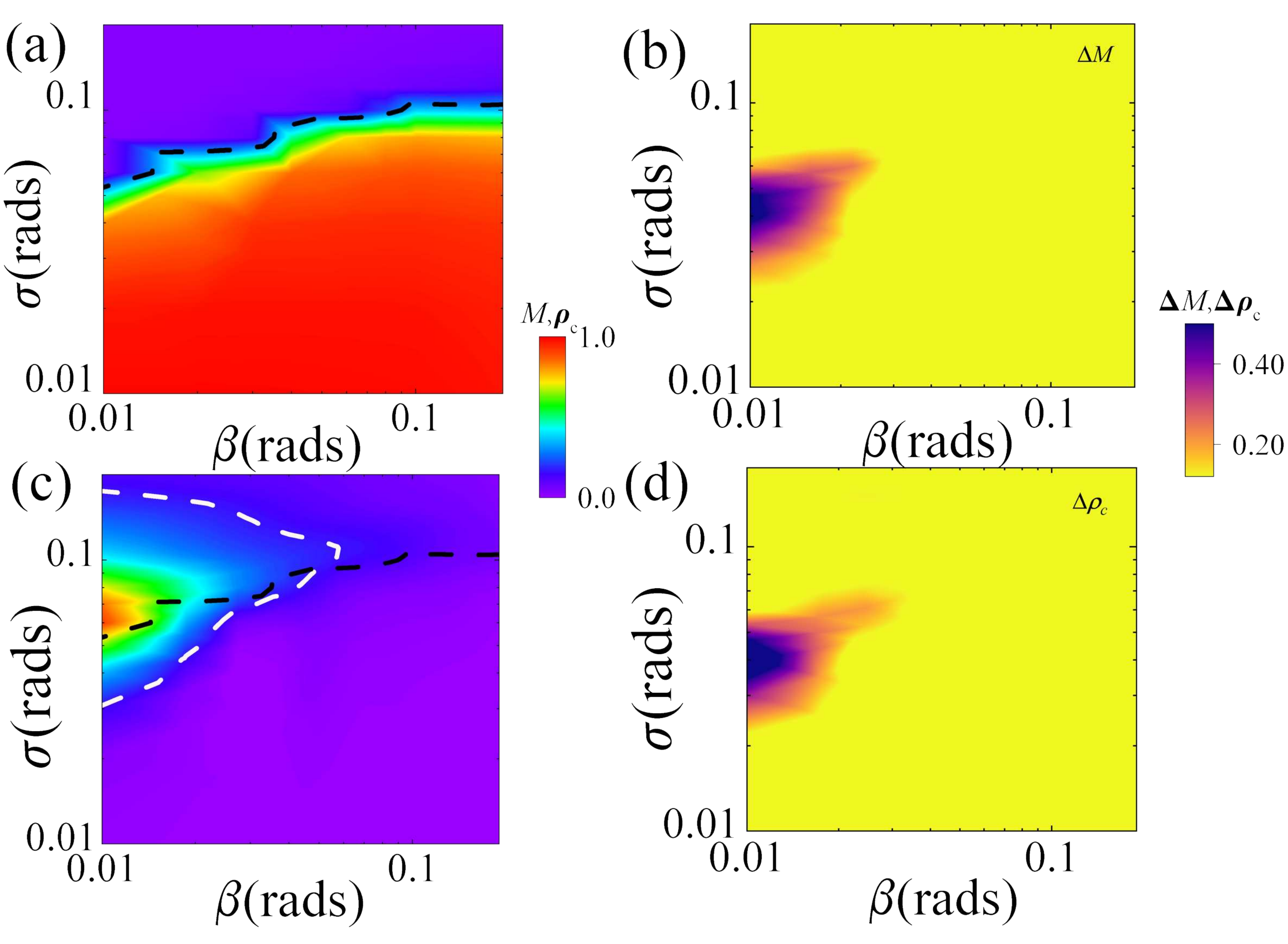}%
		\caption{(color online) Phase diagrams of steady states starting from highly alignment condition. (a) Phase diagram of $M$. (b) Phase diagram of $\rho_c$. (c) The difference of $M$ between Fig.\,\ref{thre_domains_05_25_normal}(a) and Fig.\,\ref{deltam_and_delta_roc}(a). (d) The difference of $ \rho_c$ between Fig.\,\ref{thre_domains_05_25_normal}(b) and Fig.\,\ref{deltam_and_delta_roc}(b).}
		\label{deltam_and_delta_roc}
\end{figure}. 

 In equilibrium systems, the phase transition is usually independent of the initial state. But this argument may not hold for non-equilibrium systems.  So in this session, we explore the influence of initial states such as velocity distribution on the phase diagram, evolutional process and the distribution of order parameter in the steady state.
\subsection{The effect of  initial states on phase diagrams}
We denote the value of  initial orientational order parameter as $M_0$. We change initial states of the system from the homogeneous state ($M_0$=0) to the highly flocking state ($M_0$=1). We thought no matter starting from a random configuration or from perfect alignment configuration, the final steady state would be the same. However, we strikingly find that this is not always the case.  Fig.\,\ref{deltam_and_delta_roc}(a) and (b) represent phase diagrams of $M$ and $\rho_c$ in the steady state under the initial state from highly alignment condition. Comparing phase diagram of $M$ in  Fig.\,\ref{thre_domains_05_25_normal}(a) with Fig.\,\ref{deltam_and_delta_roc}(a), we can see there are differences between them.  The phase separation line obviously shifts when $\alpha$ is relative small. And the flocking domain is broaden in Fig.\,\ref{thre_domains_05_25_normal}(a) than it in Fig.\,\ref{deltam_and_delta_roc}(a).   As for the phase diagram of $\rho_c$, on the contrary,  the clustering domain in Fig.\,\ref{deltam_and_delta_roc}(b) becomes smaller than the domain in  Fig.\,\ref{thre_domains_05_25_normal}(b).

Next we quantitively characterize the difference between these diagrams. Fig.\,\ref{deltam_and_delta_roc}(c) shows the difference between orientational order parameters in the steady states under two different initial states mentioned above. The differences are concentrated in the dark area.  Likewise,  we present the difference between $\rho_c$ under the same condition in Fig.\,\ref{deltam_and_delta_roc}(d). We observe that they are quite similar to each other,  basically matching up in the same domain. 
Then we define the correlation $C_{c o r}$ as

    \begin{equation}
      C_{c o r}=\frac{\int \Delta \rho_{c} \Delta M d \alpha d \sigma}{\sqrt{\int (\Delta \rho_c)^{2} d \alpha d \sigma \int(\Delta M)^{2} d \alpha d \sigma}}
	\end{equation}
By calculating the correlation between the two phase diagrams, we find that the correlation is about 60.43$\%$, which suggests a strong correlation between flocking and clustering. 

\begin{figure}[t]
\includegraphics[width=1.02\linewidth]{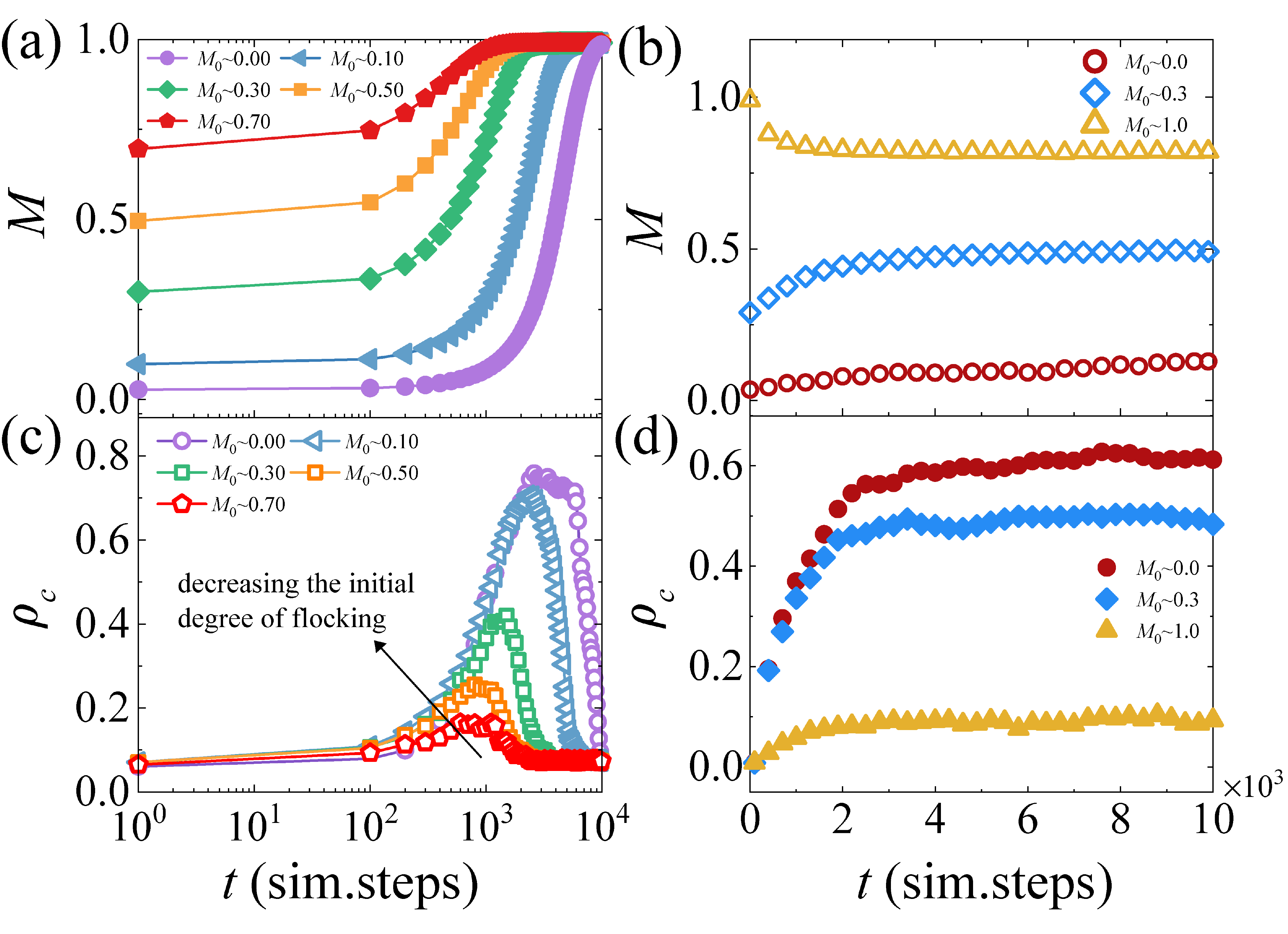}%
		\caption{(color online) Time evolution of $M$ and $\rho_c$ from different initial states with parameters $\alpha=0.01$,$\sigma=0.01$ (left column) and $\alpha=0.01$,$\sigma=0.04$ (right column).
		}
		\label{diff-M0-07-02}
\end{figure}. 

\subsection{The influence of initial states on the evolutionary process and steady states}
In order to study how initial states affect the evolution of system, we choose two sets of parameters in different regions. One set of parameters is $\alpha=0.01$ and $\sigma=0.01$, which belongs to the region that the steady state is not sensitive to the initial state. And the other set is $\alpha=0.01$ and $\sigma=0.04$, which belongs to the region that the steady state is sensitive to the initial state. Fig.\,\ref{diff-M0-07-02}(a) and (b) are the evolution of $M$ with time under different initial states, and Fig.\,\ref{diff-M0-07-02}(c) and (d) are $\rho_c$ likewise. 
Fig.\,\ref{diff-M0-07-02}(a) and (c) reveal that with the increase of $M$, $\rho_c$ accumulates gradually. When $M$ of the system is small, the cluster promotes $M$ and increases the growth of $M$. With $M$ getting larger, $\rho_c$ also increases. This is a positive correlation between $M$ and $\rho_c$. When $M$ $\approx$ 0.5, the whole system is in the state of the strongest clustering state. And then as $M$ continues to increase, $\rho_c$ begins to disintegrate. In Fig.\,\ref{diff-M0-07-02}(a), we observe that initial states influence the growth of $M$, that is to say, the growth rate of $M$ decreases from disorder to highly alignment states under the same time. The behavior of $dM/dt$ synchronizes with $\rho_c$ which is shown in Fig.\,\ref{diff-M0-07-02}(c). Meanwhile, the ability of clustering decreases with the increase of $M_0$. So we can understand that it is initial states that affect the evolution of the system, reflecting on the change of cluster formation and local alignment. 

In the case above, the initial state affects the evolution of the system, but does not affect steady states at all. Further, we explore the other case $\alpha=0.01, \sigma=0.04$ which is in the domain sensitive to initial states. 
From Fig.\,\ref{diff-M0-07-02}(b) and (d), we obviously observe that the system finally stabilizes at different amplitudes with different initial states  when the system reaches steady states. We realized that different initial states of the system lead to distinctly non-equilibrium steady states in the certain critical domain.


\begin{figure}[t]
\includegraphics[width=1.0\linewidth]{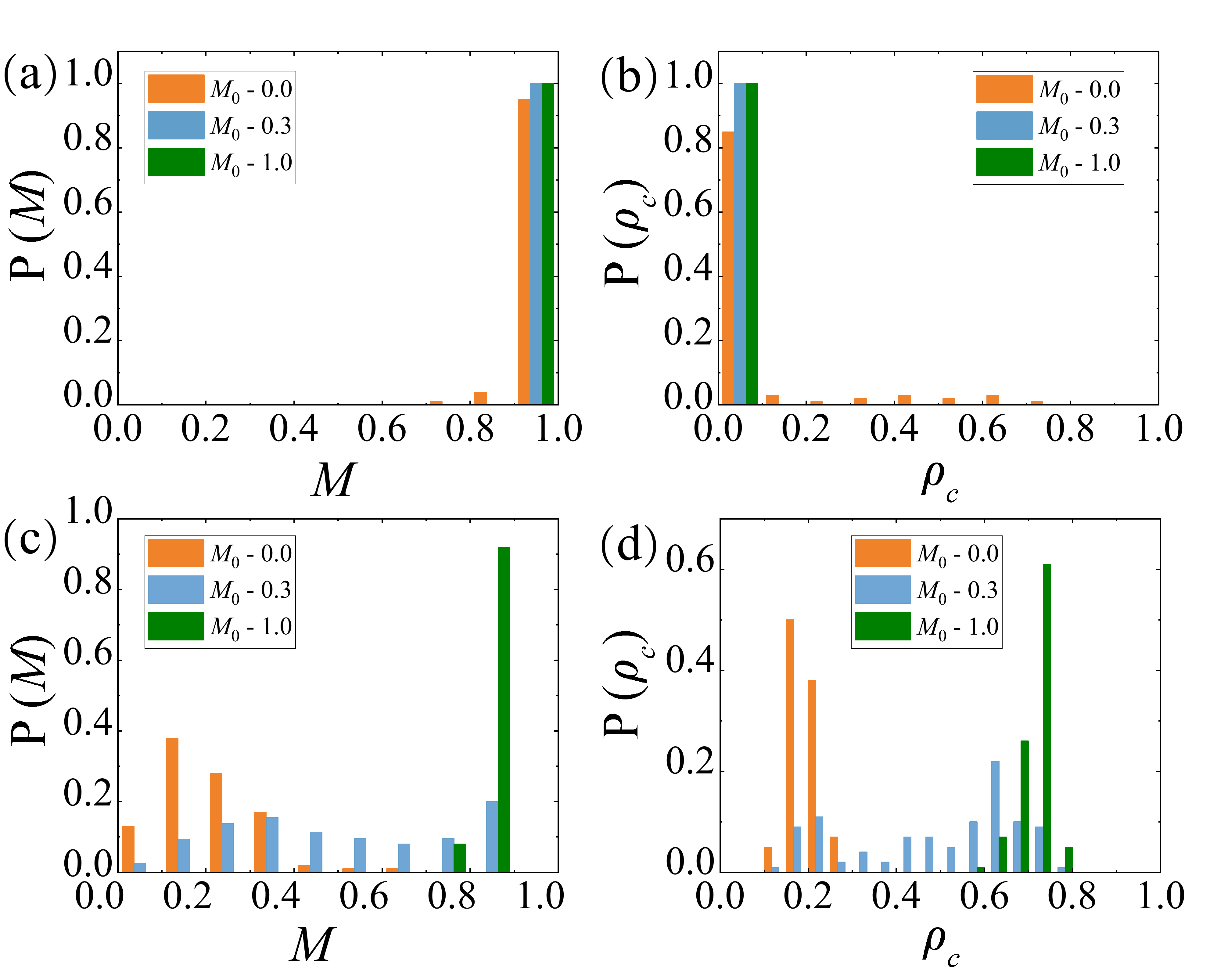}%
		\caption{(color online) Distribution of order parameters in steady states. (a) and (b) are the distributions of $M$ and $\rho_c$ with parameters $\alpha=0.01$,$\sigma=0.01$ respectively. (c) and (d) are the distributions of $M$ and $\rho_c$ with parameters $\alpha=0.01$,$\sigma=0.04$.
		}
		\label{distribution}
\end{figure} 
\subsection{Distribution of order parameters in the steady-state}

Whether the final state of the system is completely maintained in a middle-flocking state, or a result of ensemble average is the key to understanding this non-equilibrium self-organization phenomenon. Thus we explore the distribution of $M$ and $\rho_c$ in the steady states evolved from different initial states. As for the steady states in the domain which is insensitive to initial states, Fig.\,\ref{distribution}(a) and (b) reveal that the distribution of $M$ and $\rho_c$ in the steady states is totally located at around 1 and 0, respectively. However, in the domain which is sensitive to  initial states, according to the distributions shown in Fig.\,\ref{distribution}(c) and (d), we see that it is possible to move from the bistable regime into the only disordered regime or ordered regime by decreasing or increasing $M_0$. We would like to emphasize that such behaviors only appear in the specific domain of the phase diagrams. 

So far, we bring out the existence of a metastability regime which is affected by initial states. However, we still need a understanding of phase-ordering kinetics because the absence of a well-defined notion of temperature and free energy of these systems far from equilibrium makes us difficult to get insights from a theoretical point of view.

\section{CONCLUSION}
To conclude, we have investigated the influence of initial states on the evolution process and steady state of the active system based on a model with active reorientation. The system displays rich phase behaviors including global flocking, clustering and disordered phases. We observe an interesting interplay between flocking and clustering. We notice that the boundary of flocking phase and clustering phase rely on the choice of initial states. Although clustering is like a high nucleation barrier that prevents the system from escaping from metastable states~\cite{Softmatter81}, the increasing of the order degree of initial states would suppress clustering during the process of self-organization evolution. In the meantime, we find that there is an unstable region in our model, in which the system will relax to different steady states due to different initial conditions. This is consistent with the ABP with alignment theory proposed by R. Großmann et al. However, the difference between our model and theirs is that there is no clear alignment, which also shows this similar nature. In addition, under the competition between external noise and active angular fluctuations, our system can bypass some states that should be experienced under different initial conditions, that is, the dynamic path, it evolves into different stable states. In the work by Chvykov et al~\cite{scienceabc6}, the possible future state of active robots can be controlled by adjusting the initial state of the system. Our work provide a theoretical evidence for the experiment mentioned above. Since dynamic path can influence evolution behaviors of self-organizing system, this means that active matter or active robots may be controlled to allow the system to grow towards an expected state from the perspective of regulation. This finding has a wide range of applications in improving designs of collective migration and navigation strategies.

\begin{acknowledgments}
We wish to acknowledge Xinliang Xu offering suggestions and encouragement. This research was supported by the National Natural Science Foundation of China (Grants No. 11975050, No. 11735005 and No. 11904320). We also acknowledge computational support from the Beijing Computational Science Research Center.
\end{acknowledgments}

{}

\bibliography{apssamp}

\end{document}